# Enhancing Diabetic Retinopathy Detection with CNN-Based Models: A Comparative Study of UNET and Stacked UNET Architectures

[1] S. Navaneetha Krishnan, [2]Ameya Uppina, [3]Talluri Krishna Sai Teja, [4]Nikhil N Iyer, [5]Joe Dhanith P R

School of Computer Science and Engineering, Vellore Institute of Technology Chennai Campus, India

Email: [1]krishnannavaneetha467@gmail.com, [2]ameyauppina2003@gmail.com, [3]teja05443@gmail.com, [4]nikhiliyer95@gmail.com, [5]joedhanith.pr@vit.ac.in

**Abstract:** Diabetic Retinopathy (DR) is a severe complication of diabetes. Damaged or abnormal blood vessels can cause loss of vision. The need for massive screening of a large population of diabetic patients has generated an interest in a computer-aided fully automatic diagnosis of DR. In the realm of Deep learning frameworks, particularly convolutional neural networks (CNNs), have shown great interest and promise in detecting DR by analyzing retinal images. However, several challenges have been faced in the application of deep learning in this domain. High-quality, annotated datasets are scarce, and the variations in image quality and class imbalances pose significant hurdles in developing a dependable model. In this paper, we demonstrate the proficiency of two Convolutional Neural Networks (CNNs) based models – UNET and Stacked UNET utilizing the APTOS (Asia Pacific Tele-Ophthalmology Society) Dataset. This system achieves an accuracy of 92.81% for the UNET and 93.32% for the stacked UNET architecture. The architecture classifies the images into five categories ranging from 0 to 4, where 0 is no DR and 4 is proliferative DR.

## I. INTRODUCTION

With increased diabetes among individuals in developed nations, Diabetic Retinopathy (DR) acts as a major cause of blindness. It represents a severe complication of Diabetes Mellitus (DM), resulting in significant vision impairment due to damage to the retina. According to various studies, an estimated 420 million people worldwide are diagnosed with diabetes[1][2], with the incidence of the disease having doubled over the past 30 years. Projections indicate that this number will continue to rise. Around one-third of diabetic patients are expected to develop DR, with 40% of type II diabetes patients and 86% of type I diabetes patients in the US showing signs of DR. Additionally, rural areas in China report DR rates as high as 43%[3]. Regular eye examinations for the early detection of DR are recommended by major ophthalmological organizations like American Academy of Ophthalmology, and The International Council of Ophthalmology[4].

Vision loss associated with $diabetic\ retinopathy\ (DR)$ occurs gradually as the condition progresses. DR is commonly categorized into two primary stages: $non-proliferative\ diabetic\ retinopathy\ (NPDR)$ and $proliferative\ diabetic\ retinopathy\ (PDR)$, the latter being marked by neovascularization or hemorrhages in the vitreous or pre-retinal areas[4]. The American Academy of Ophthalmology classifies DR into five stages based on severity. The first stage, "no apparent retinopathy," reflects a normal retinal examination. The next three stages—mild, moderate, and severe NPDR—represent escalating retinal damage. Mild NPDR involves the presence of microaneurysms without significant additional damage, while severe NPDR is characterized by 20 or more hemorrhages per retinal quadrant, followed by intraretinal microvascular abnormalities and venous beading in at least two quadrants. Moderate NPDR falls between mild and severe stages. The final stage, PDR, involves the growth of new blood vessels on the retina (retinal neovascularization) or bleeding into the vitreous humor (vitreous hemorrhage) or between the retina and vitreous membrane (preretinal hemorrhage)[5]. Retina specialists often use fluorescein angiography to assess the severity of DR and identify areas of macular damage[6][7].

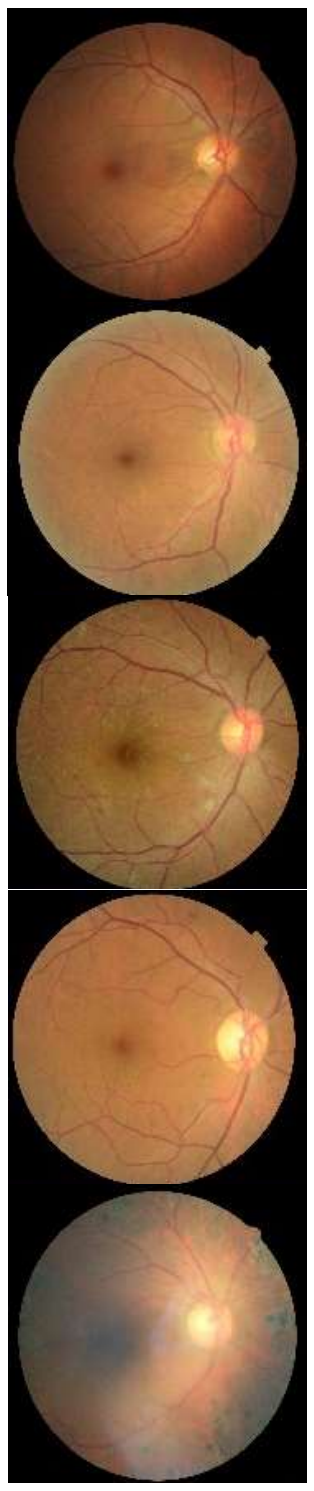

*Figure 1: Stages of DR. (From Top): No DR, Mild DR, Moderate DR, Severe DR, and Proliferative DR.*

Common symptoms of DR include blurry vision, the appearance of floaters or flashes, and vision loss[8]. DR affects the macula due to metabolic changes in the retinal blood vessels, which result in abnormal blood flow and leakage of blood components onto the retina. This leakage causes the retinal tissue to swell, leading to clouded or blurred vision. DR typically affects both eyes and, if left untreated, can lead to diabetic maculopathy—a form of blindness resulting from long-term diabetes without proper management[9]. If DR goes undiagnosed or untreated, its progressive nature can worsen vision, ultimately leading to irreversible blindness. Over time, retinal lesions form as a result of ruptured retinal blood vessels, including $microaneurysms\ (MAs)$, $hemorrhages\ (HEs)$, $exudates\ (EXs)$, $cotton\ wool\ spots\ (CWSs)$, and other abnormalities like fibrotic bands, $intra-retinal\ microvascular\ abnormalities\ (IRMAs)$, $neovascularization\ on\ the\ disc\ (NVD)$, and elsewhere (NVE), as well as tractional bands[10][11]. These retinal lesions affect the fundus, and the back of the eye, and their timely detection is critical for identifying the various stages of DR[12].

Early detection of DR involves closely monitoring key retinal structures, including the optic disc (OD), retinal blood vessels (RBVs), fovea, and macula. This is often achieved through pupil dilation facilitated by medically approved contrast agents injected into the retina. Techniques such as FA (fluorescein angiography) and mydriatic fundus photography are used to capture detailed images of the fundus, which are then analyzed to detect DR[13] in its early stages. Traditional diagnostic methods[14] include bio-microscopy, fundus imaging, retinal thickness analysis, $scanning\ laser\ ophthalmoscopy\ (SLO)$, adaptive optics, $retinal\ oximetry, optical\ coherence\ tomography\ (OCT)$, OCT angiography, and Doppler OCT. However, these approaches can be labor-intensive, time-consuming, and prone to error, often requiring highly trained specialists, which can be impractical given the shortage of ophthalmologists, particularly in regions like India where the ratio of ophthalmologists to patients is 1:10,000[15]. This underscores the need for an automated, intelligent detection system capable of conducting early screenings for DR. Developing such systems, powered by large datasets of fundus images, could significantly improve early diagnosis, potentially preserving the vision of diabetic patients through timely intervention.

With the rise of machine learning, numerous computer-assisted systems have been developed for detecting DR using algorithms such as SVMs, Neural Networks, and Decision Trees. However, traditional machine learning algorithms struggle when dealing with large-scale, high-dimensional, and complex data like medical images. These methods often require significant computational resources and lack the flexibility and domain-specific insights needed for accurate data representation. This led to the emergence of deep learning, a more advanced branch of machine learning, which is better suited for handling complex tasks, extracting deep features, and processing high-dimensional data. Deep learning frameworks are highly scalable, capable of learning domain-specific knowledge, and can make reliable decisions, outperforming traditional models. Techniques such as Convolutional Neural Networks (CNN)[16][17], Deep Convolutional Neural Networks (DCNN), and other architectures like AlexNet, VGG-16, VGG-19[18], GoogLeNet, ResNet, DenseNet, and Inception have been employed for deep feature extraction and image classification. Additional methods such as Autoencoders, Restricted Boltzmann Machines (RBMs), LSTM networks, and Deep Reinforcement Learning (DRL) have also been explored [19]. Despite these advances, challenges remain, particularly due to the scarcity of high-quality annotated datasets, variations in image quality, and class imbalances that complicate model development.

This paper presents a comparison of two CNN-based architectures—UNET and Stacked UNET—for the early detection of DR using the APTOS 2019 dataset. APTOS dataset comprises 3,662 retinal images collected from participants in rural India, with data organized by the Aravind Eye Hospital. These fundus images were gathered under diverse conditions over an extended period. A team of trained ophthalmologists later reviewed and labeled the samples following the guidelines of International Clinical DR Disease Severity Scale (ICDRSS). Based on this scale, the images were categorized into five stages: no DR, mild DR, moderate DR, severe DR, and proliferative DR. UNET was chosen for its capability to use convolutional layers and max pooling operations to extract distinct features and patterns from the retinal images.

The original images, sized at 512x512x3 (RGB), were resized to 256x256x1, converting them to grayscale. To expand the dataset, horizontal flips of the images were generated, retaining the same label as the original. The input layer for the model thus became 256x256x1. In a CNN, the convolutional layer comprises multiple filters that are applied to the input through convolution operations. Max pooling was then employed, using a smaller matrix or kernel that scans the convolutional layer. The application of 2D convolution and max pooling layers multiple times allows for the expansion of image dimensions. The number of layers and pooling operations varies depending on the architecture, ultimately forming dense layers where neurons are fully connected, which influences the final output size and shape.

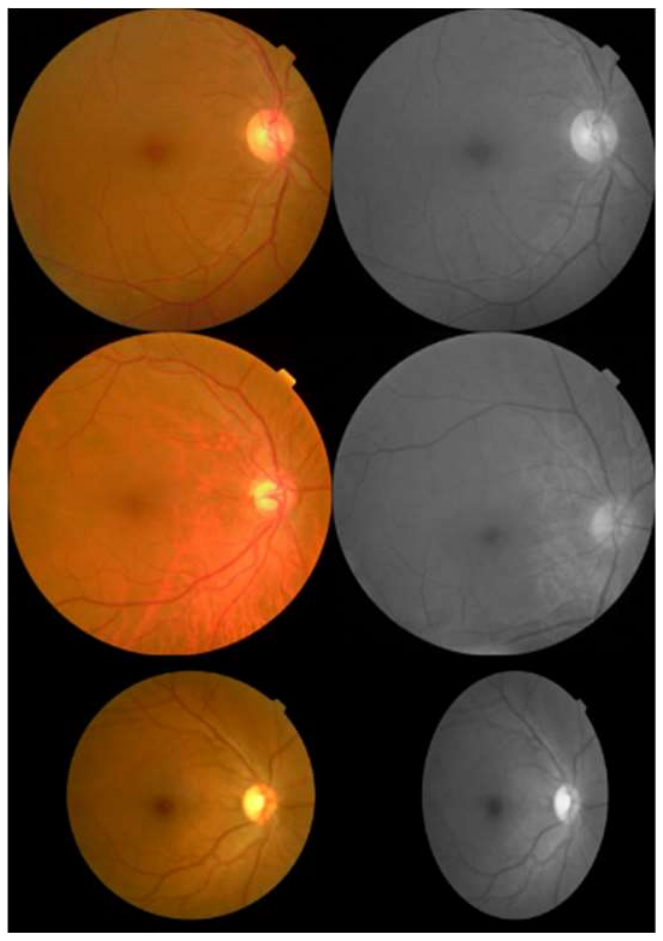

*Figure 2:*
*Left Column: Raw Images*
*Right Column: Preprocessed Images*

## II. PROPOSED METHODOLOGY

The proposed UNET architecture aims to enhance pixel-patch-based localization by utilizing Upsampling instead of pooling, thus increasing its applicability in the biomedical domain. This architecture integrates high-resolution features from the contracting path with Upsampled outputs, ensuring precise segmentations without fully connected layers. We are working with 256 x 256 x 1 sized images which have been converted to black and white.

The UNET architecture is divided into two parts: Downsampling and Upsampling. In the Downsampling sector, each block is called an encoder block, consisting of two convolution layers and one max pooling layer. Each layer gets the input from the output of the previous layer, decreasing the size of the image being generated and effectively zooming in on the required patches in the image. There are three encoder blocks in the Downsampling sector. The first layer employs 64 filters followed by 128 and 256 in subsequent layers. The size of the image decreases as follows: an input of size 256 x 256 is reduced to 128 x 128 for the second layer, followed by a 64 x 64 image which is further reduced to 32 x 32. The features generated at each layer are stored for integration with the outputs generated during the Upsampling portion of the UNET. The images generated are then passed through 2 other convolution layers before proceeding to the Upsampling section of the UNET.

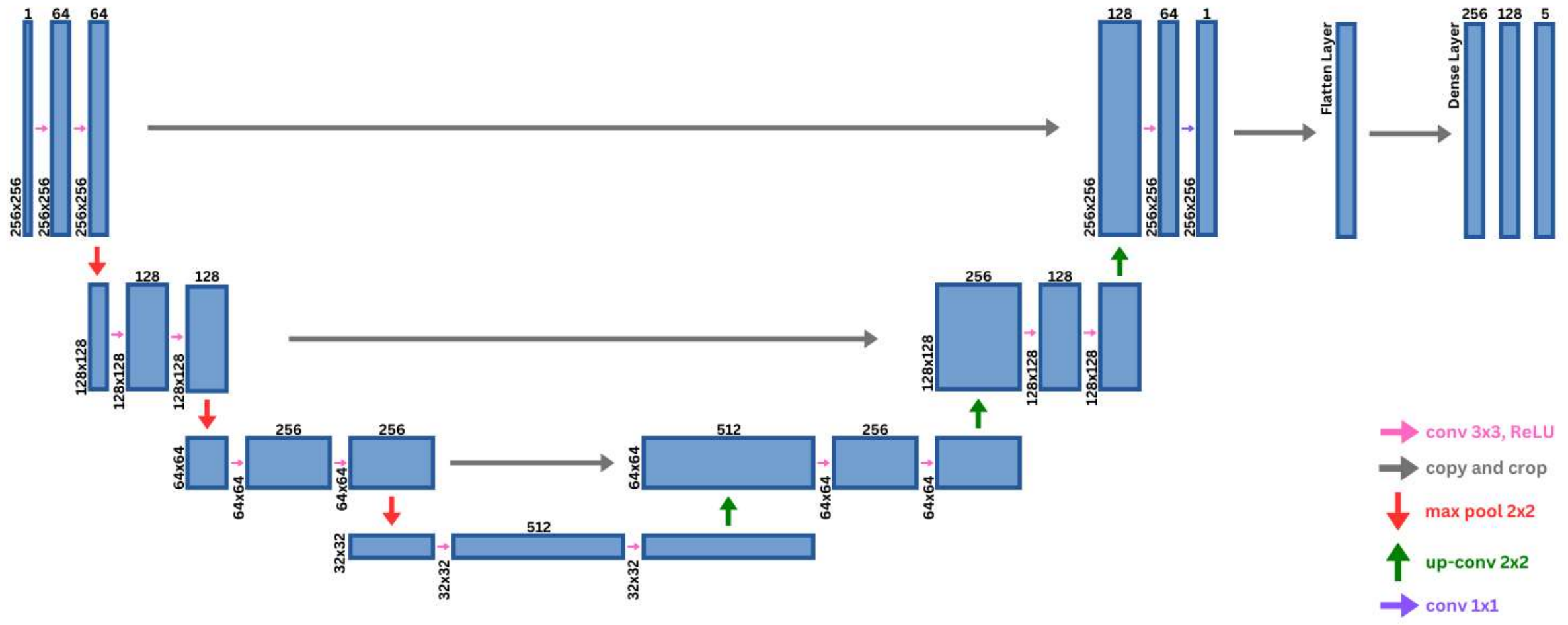

*Figure 3: Model Architecture of UNET*

In the Upsampling sector, each block is called a decoder block, consisting of a Convolution 2D Transposition layer, which reconstructs spatial information using the features generated from the previous sections followed by a concatenate layer to combines the information from the transposition layer with the features from its respective sector in the Downsampling section, and a Convolution layer. The number of filters for each block follows the order of 256, 128, and 64. The output sizes of each layer mirror those of the Downsampling sector, only in reverse order. Therefore, the final image size of the UNet architecture is the same as the input size, i.e., 256 x 256 in grayscale. The output is then passed onto a flattened layer which converts the entire image into a tensor. This tensor is then used as the input for an array of 3 Dense layers of sizes 256,128 and 5 in order, which classify the image into the required number of classes, in this case, five.

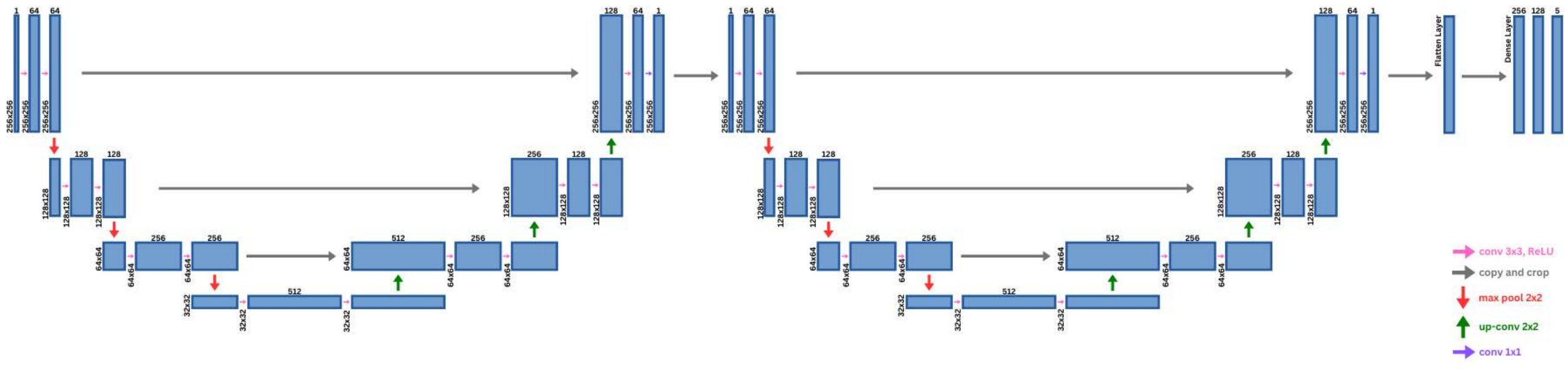

*Figure 4: Model Architecture of Stacked UNET*

For 2 stack UNET architecture, the 256 x 256 output of the Upsampling section is sent as input to the Downsampling part of the second UNET layer. The 256 x 256 output of the 2nd layer is then used for prediction. It is passed onto Flatten and Dense layers. The flattening layer converts the entire image into a tensor, which is given as input to the dense layer. This layer then takes the tensor and uses it for the classification into the required number of classes, in this case, five. The Conv2D layer (2D convolutional layer) is a basic building block in convolutional neural networks used for image analysis and other tasks.

Convolution operations are performed on the preprocessed set of images using kernels, which are a set of learnable filters. The filters slide over the input data, computing element-wise products and summing them up to produce feature maps. The Conv2D layer has parameters like the number of filters, filter size (kernel size), padding, and activation function. In our model, we use a filter size of 3 x 3 and padding as 'same,' which maintains the input size.

## III. RESULTS

The UNET model shows excellent training performance with a low loss of 0.0167 and an accuracy of 99.35%, along with precision, recall, and F1 scores all near 99%, indicating strong performance on the training set. However, its validation performance drops, with a higher loss of 0.3175 and an accuracy of 92.81%. While precision remains high at 98.57%, recall falls to 92.57%, leading to a decrease in the F1 score to 91.45%, suggesting overfitting. In comparison, the Stacked UNET model performs similarly on the training set with a slightly lower loss of 0.0117 and an accuracy of 99.28%, maintaining high precision, recall, and F1 scores around 99%. Its validation performance shows improvement over the standard UNET, with a lower loss of 0.2700 and a higher accuracy of 93.32%. The precision, recall, and F1 scores are also better, with an F1 score of 92.85%, indicating that the stacked architecture helps reduce overfitting and improves generalization. Overall, the Stacked UNET outperforms the base model in terms of validation metrics, making it more robust for unseen data.

The Stacked UNET's deeper architecture allows it to capture more complex features, improving its ability to generalize to new data. This architecture also likely improves spatial resolution during upsampling, enhancing performance in tasks requiring precise localization. Additionally, stacking helps the model mitigate overfitting by distributing learning across multiple layers. These advantages make the Stacked UNET more reliable for real-world applications.

***Table 1: Results of UNET and Stacked-UNET***

| Model | | Loss | Accuracy | AUC | Precision | Recall |
|---|---|---|---|---|---|---|
| **UNET** | **Training** | 0.0167 | 99.35 | 100 | 99.04 | 99.02 |
| | **Validation** | 0.3175 | 92.81 | 98.57 | 92.57 | 91.45 |
| **Stacked UNET** | **Training** | 0.0117 | 99.28 | 100 | 99.29 | 99.28 |
| | **Validation** | 0.2700 | 93.32 | 98.80 | 93.80 | 92.85 |

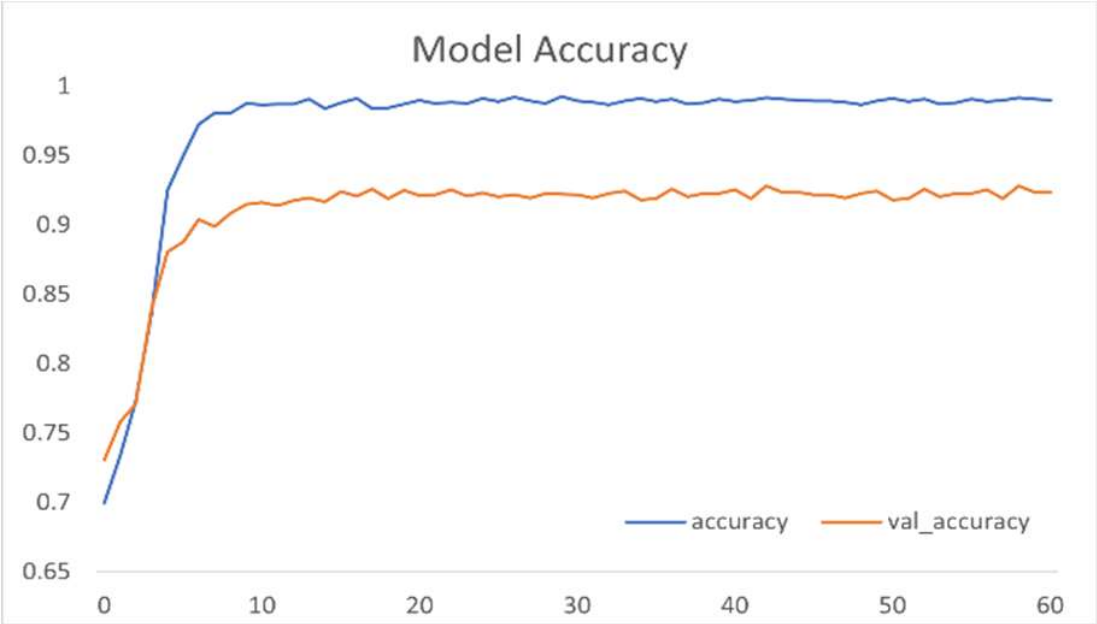

*Figure 5: Accuracy of UNET*

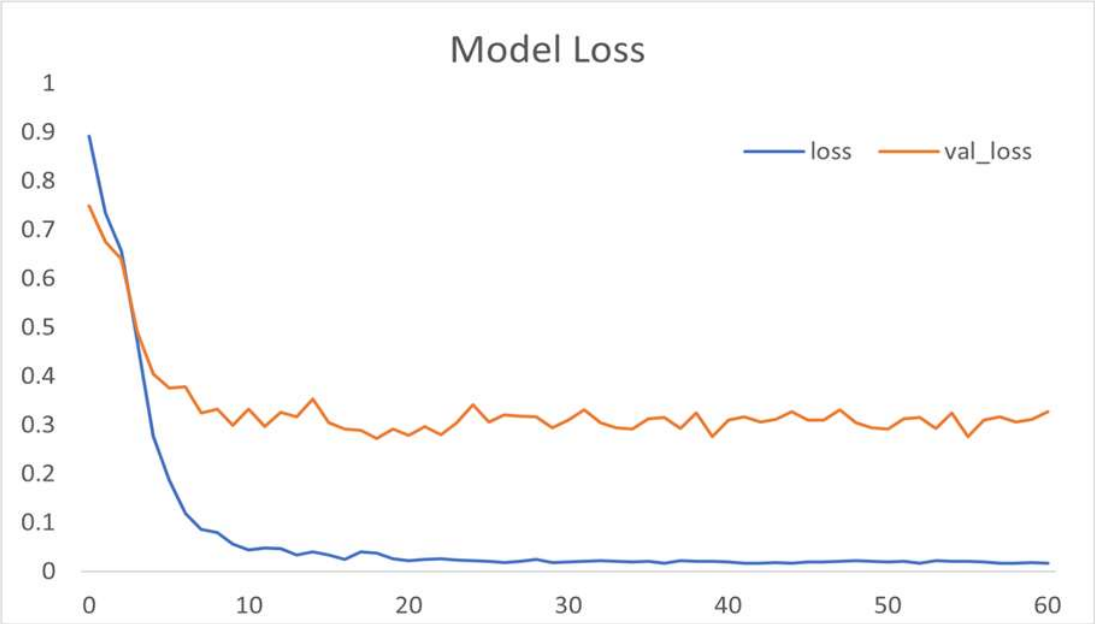

*Figure 7: Loss of UNET*

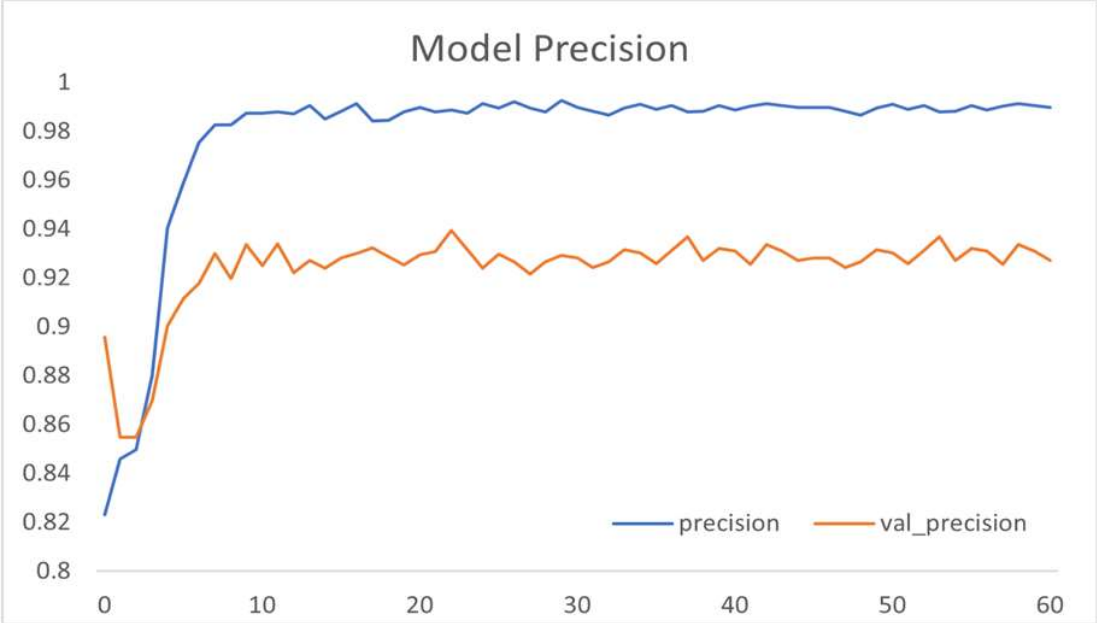

*Figure 6: Precision of UNET*

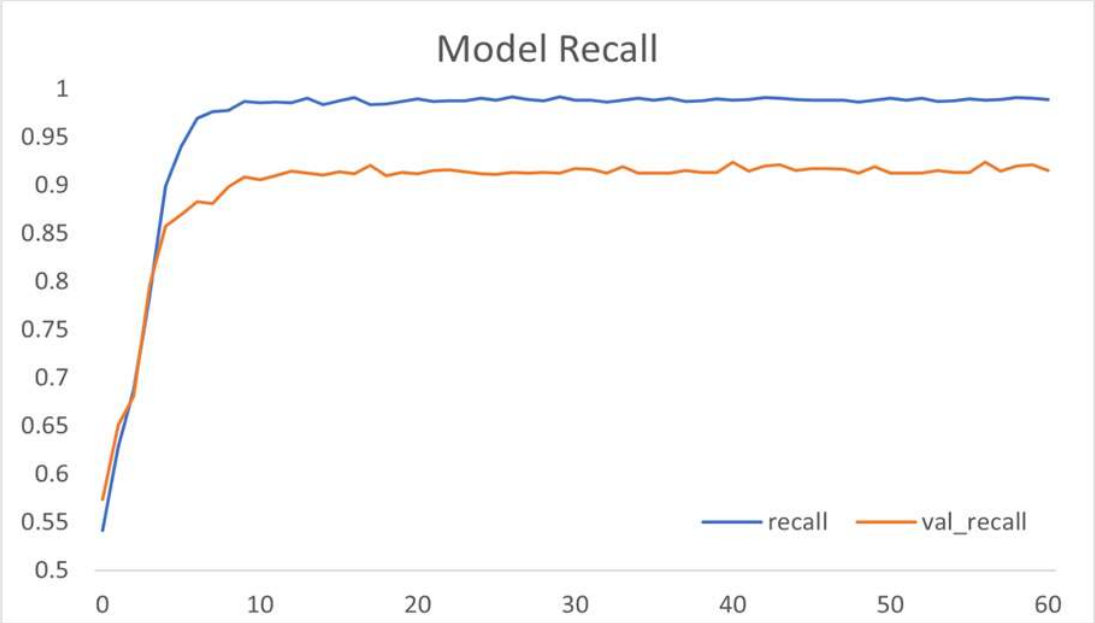

*Figure 8: Recall of UNET*

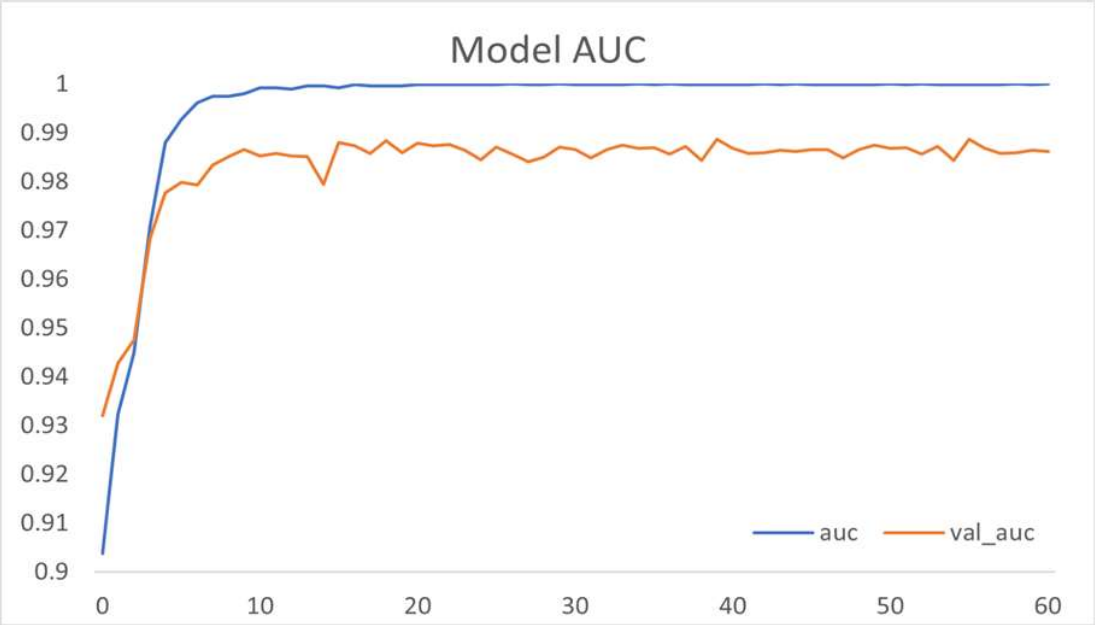

*Figure 9: AUC of UNET*

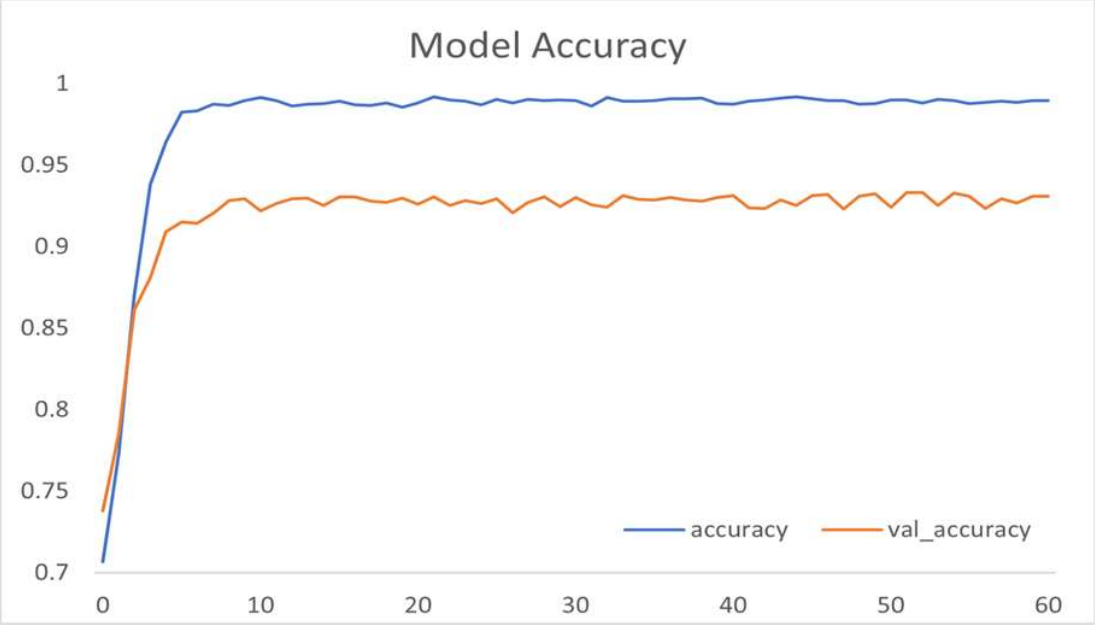

*Figure 10: Accuracy of Stacked UNET*

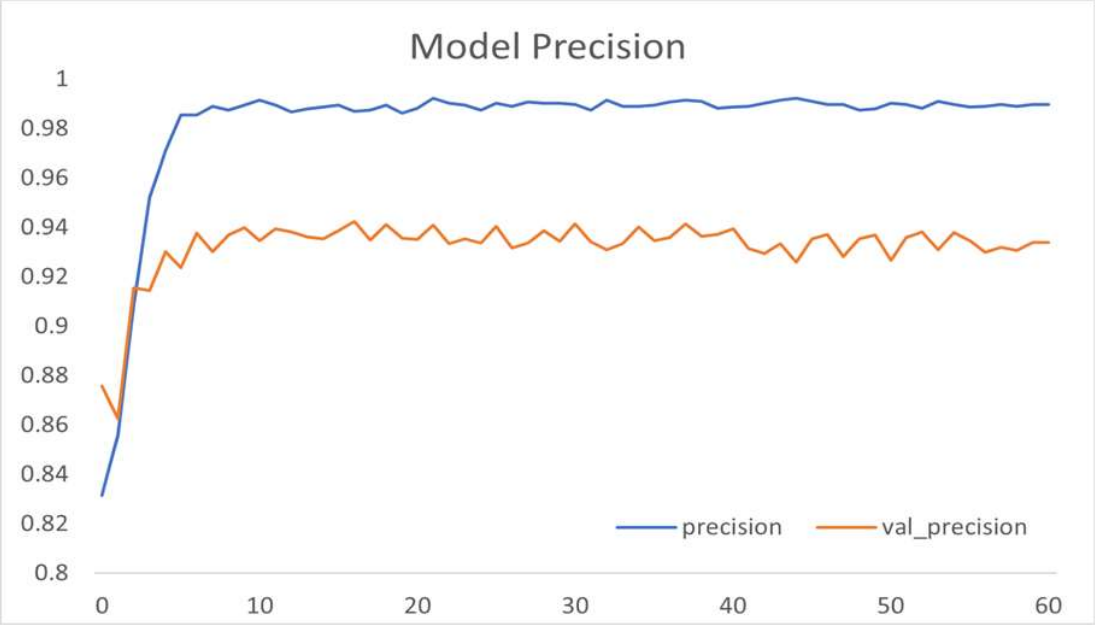

*Figure 11: Precision of Stacked UNET*

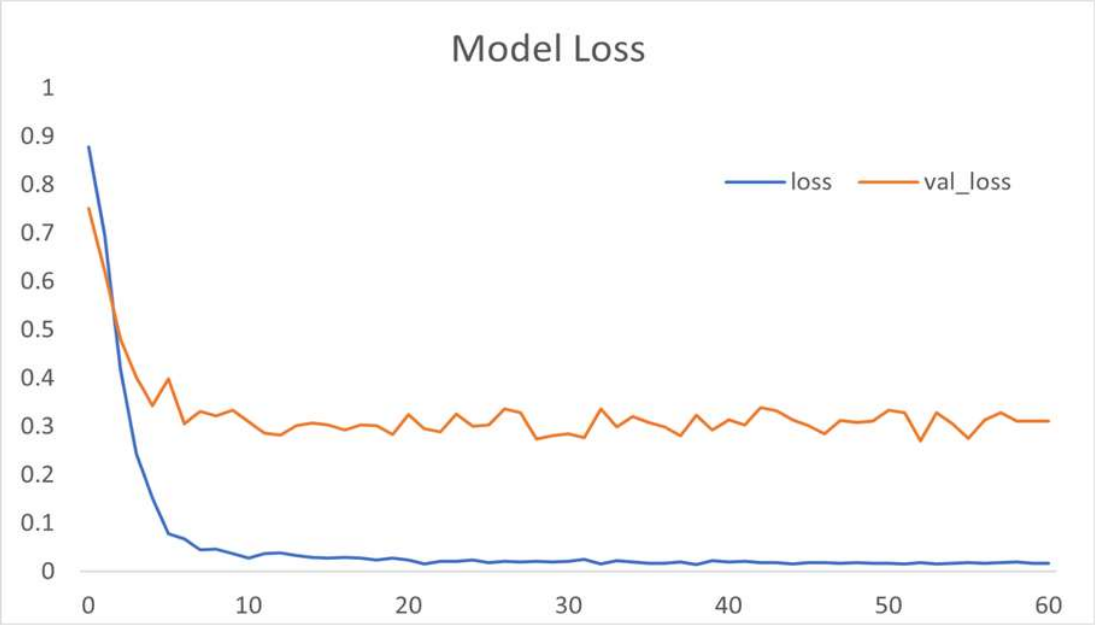

*Figure 12: Loss of Stacked UNET*

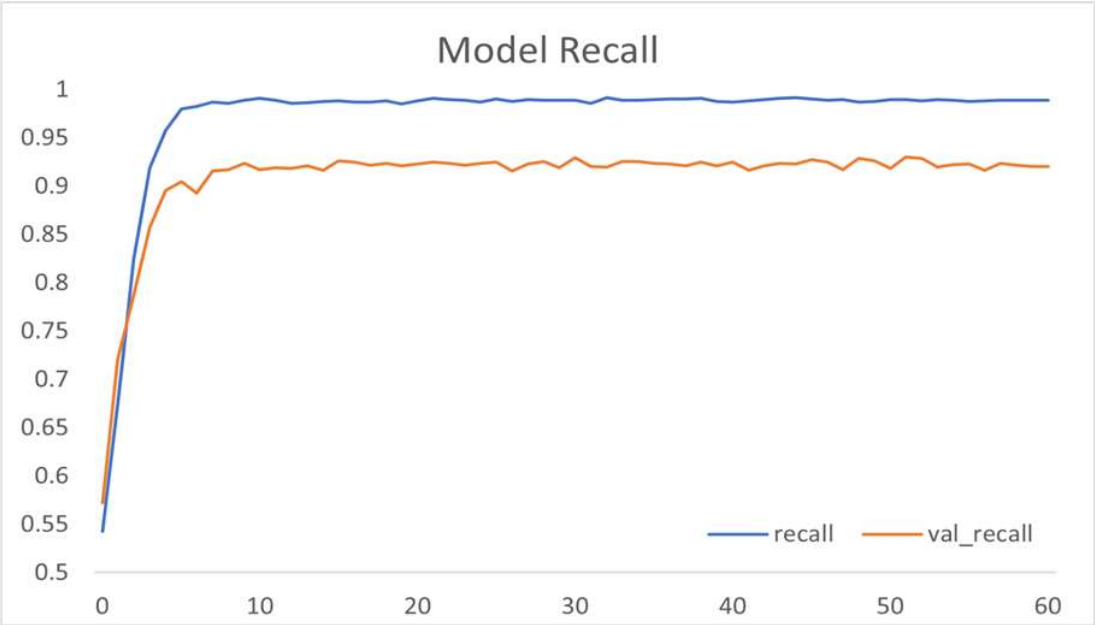

*Figure 13: Recall of Stacked UNET*

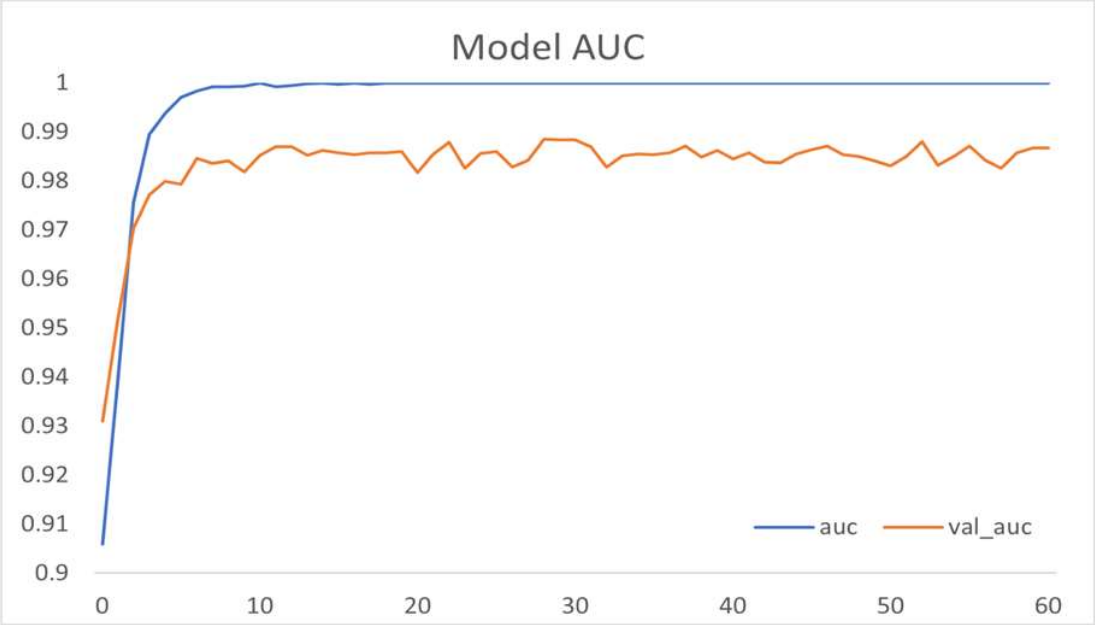

*Figure 14: AUC of Stacked UNET*

## IV. CONCLUSION

In this paper, we presented a comparative analysis of two CNN-based architectures, UNET and Stacked UNET, for the automatic diagnosis of DR using the APTOS dataset. Both models demonstrated the potential for detecting DR across varying stages, achieving notable accuracy in classifying retinal images. The UNET model achieved an accuracy of 92.81% and Stacked UNET achieved an accuracy of 93.92%, indicating promising performance despite challenges such as dataset quality, image variations, and class imbalances. The Stacked UNET, with its deeper architecture, showed improved performance, underscoring the benefits of more complex models for enhancing image classification accuracy. While these results highlight the feasibility of CNN-based models in DR diagnosis, future work should focus on further improving model robustness by addressing dataset limitations, applying data augmentation techniques, and refining model architectures to boost sensitivity and generalization for clinical applications.